\documentclass[conference]{IEEEtran}
\usepackage{amsthm,amsmath,amssymb}

\usepackage{mathrsfs}

\usepackage{cite}
\usepackage{graphicx}
\usepackage{times}
\usepackage{latexsym}

\usepackage[mathscr]{eucal}
\usepackage{array}
\usepackage{fancyhdr}
\usepackage{amsmath,amssymb}
\usepackage{bm} 
\usepackage{subfigure}
\usepackage{citesort}
\usepackage{multirow}

\ifCLASSINFOpdf
\else
\fi

\newcommand{\diag}{{\rm{diag}}}
\usepackage{xcolor}

\makeatother

\usepackage{amsthm}
\usepackage{amssymb}
\usepackage{graphicx}
\usepackage{subfigure}
\usepackage{tabularx}
\usepackage{color,cite}
\usepackage{url}
\usepackage{bm}
\usepackage{color}
\definecolor{c}{rgb}{1,0,0} % red
\definecolor{b}{rgb}{0,0,1} % red

\usepackage{algpseudocode}
\usepackage{algorithm}
\usepackage{makecell}
\usepackage{amsfonts}

\newtheorem{theorem}{Theorem}
\newtheorem{lemma}[theorem]{Lemma}

\usepackage{algorithm}

\usepackage{algpseudocode}
\begin{document}

%\title{Transmit and Reflective Beamforming based Secret Key Generation in Spatial Correlated Channels
%}

\title{Joint Transmit and Reflective Beamforming for RIS-assisted Secret Key Generation
}

%% author ----
\author{
	\IEEEauthorblockN{Lei Hu$^{*}$, Guyue Li$^{*\dag}$, Xuewen Qian$^{\P}$, Derrick Wing Kwan Ng$^{\Vert}$, and Aiqun Hu$^{\dag\ddag}$,}
	\IEEEauthorblockA{ $^{*}$School of Cyber Science and Engineering, Southeast University, Nanjing, 210096, China}
	\IEEEauthorblockA{ $^{\dag}$Purple Mountain Laboratories for Network and Communication Security, Nanjing, 210096, China}
	\IEEEauthorblockA{ $^{\P}$CentraleSupelec, Paris-Saclay University, Paris, France}
	\IEEEauthorblockA{ $^{\ddag}$National Mobile Communications Research Laboratory, Southeast University, Nanjing, 210096, China}
	\IEEEauthorblockA{ $^{\Vert}$School of Electrical Engineering and Telecommunications, University of New South Wales, Sydney, NSW, 2052, Australia}
%	\IEEEauthorblockA{\{lei-hu, guyuelee,aqhu@seu.edu.cn\}@seu.edu.cn, hongyiluo\_seu@163.com}
	\IEEEauthorblockA{Corresponding author: Guyue Li, Email: {guyuelee}@seu.edu.cn}
}% \S也可以用

\maketitle
\begin{abstract}
Reconfigurable intelligent surface (RIS) is a promising technique to enhance the performance of physical-layer key generation (PKG) due to its ability to smartly customize the radio environments. Existing RIS-assisted PKG methods are mainly based on the idealistic assumption of an independent and identically distributed (i.i.d.) channel model at both the transmitter and the RIS. However, the i.i.d. model is inaccurate for a typical RIS in an isotropic scattering environment. Also, neglecting the existence of channel spatial correlation would degrade the PKG performance. In this paper, we establish a general spatially correlated channel model in multi-antenna systems and propose a new PKG framework based on the transmit and the reflective beamforming at the base station (BS) and the RIS. Specifically, we derive a closed-form expression for characterizing the key generation rate (KGR) and obtain a globally optimal solution of the beamformers to maximize the KGR. Furthermore, we analyze the KGR performance difference between the one adopting the assumption of the i.i.d. model and that of the spatially correlated model. It is found that the beamforming designed for the correlated model outperforms that for the i.i.d. model while the KGR gain increases with the channel correlation. Simulation results show that compared to existing methods based on the i.i.d. fading model, our proposed method achieves about $5$ dB performance gain when the BS antenna correlation $\rho$ is $0.3$ and the RIS element spacing is half of the wavelength.
\end{abstract}

%\begin{IEEEkeywords}
%	Physical layer security, reconfigurable intelligent surface, secret key generation, passive beamforming.
%\end{IEEEkeywords}
\IEEEpeerreviewmaketitle

\section{Introduction}
The inherent broadcast nature of wireless medium is vulnerable to security breaches, attracting passive or active attacks from potential eavesdroppers \cite{li2021maximizing}.
%, such as eavesdropping, jamming, and modification, etc. \cite{Access} 
In contrast to conventional encryption schemes that experience difficulties in key distribution, physical-layer key generation (PKG) provides an alternative approach to establish symmetric keys between the legitimate parties. By exploiting the intrinsic randomness and the reciprocity of wireless channels, PKG is information-theoretically secure. 
Nevertheless, the essential premise to ensure the security of secret key in PKG is the existence of rich-scattering and dynamically varying channels. 
Unfortunately, this condition can hardly be satisfied to guarantee key generation performance in some harsh propagation scenarios, such as static and shadowed environments. As a result, there is a need for new technologies to improve PKG performance. 

Recently, the emergence of reconfigurable intelligent surface (RIS) provides a promising means to address the aforementioned problems. RIS is a programmable
and reconfigurable metasurface consisting of a large number of passive elements, which can be controlled  
collaboratively to alter the signal propagation environment. 
Since the key generation performance relies on the properties of fading channels, RIS could be the key enabler for improving the PKG. Inspired by this,  there are several studies on the design of RIS-assisted PKG systems. For example, in static environments, a RIS-induced randomness method was proposed in \cite{JiOTP} and the experiments in  \cite{Low-entropy} demonstrated its effectiveness. In addition, the authors of \cite{li2021maximizing},\cite{Ji},\cite{SPletter}  investigated the optimization of RIS beamforming in dynamic environments to further improve the key generation rate (KGR). 
Nevertheless, all of these works are based on the independent and identically distributed (i.i.d.) Rayleigh fading model for the RIS-related channels. In practice, the 
non-negligable spatial correlations exist among RIS elements due to their sub-wavelength sizes. 
More importantly, these correlations may jeopardize the PKG performance if they are not taken into account in the system design \cite{Rayleigh-fading}.  
In addition, only a single-antenna BS was considered in these works, e.g., [1]--[5] and their results are not applicable to the case of multi-antenna. 
%However, the BS in practical systems is often equipped with multiple antennas. which provides the potential to further improve the KGR . 
Indeed, the RIS-assisted PKG methods in multi-antenna spatially correlated channels are still unknown. 

%Specifically, the authors of \cite{Ji_IRS} considered a scenario with multiple eavesdroppers and optimized the RIS reflection coefficients to maximize the lower bound of secret key rate. In \cite{SPletter}, with the number of RIS units that can be turned on is limited, the authors optimized the placement of RIS elements to maximize the secret key rate. 
%Nevertheless, the spatial correlation between BS antennas and RIS elements, which appears in practice and affects the PKG performance [sKG in ], are both not taken into account. 
% We note that in this work, 
%the spatial correlation between antennas are ignored,
%% the length of pilots scales with the number of antennas at base station, which brings large pilot overhead especially in massive MIMO network. 
%but field measurements show that the channels are often spatially correlated in some scenarios, such as outdoor environments \cite{chizhik2003multiple}. 
%Furthermore, all of existing RIS-aided PKG works are based on the independent and identically distributed (i.i.d.) Rayleigh fading model, which is not physically appearing when employing an RIS in isotropic scattering environments \cite{Rayleigh-fading}. 

To fill this gap, this paper investigates a RIS-asisted PKG method in a multi-antenna system with the consideration of the spatial correlation between the BS and the RIS. The main contributions of this paper are listed as follows: 
\begin{itemize}
	\item We propose a novel transmit and reflective beamforming based RIS-assisted PKG framework in spatially correlated channels. We formulate the design of beamforming as an optimization problem by deriving the closed-form KGR expression.
	
	%	Notably, we take the channel correlation at BS antennas and RIS elements into account, which are more practical than the existing RIS-assisted PKG model. 
	
	%给出结论。先总括，找到最优的beamforming变量，然后讲
	\item We design the globally optimal transmit and reflective beamforming vector by decomposing the optimization problem into two sub-problems and optimizing them separately. 
	Also, our analysis shows that the optimal beamforming outperforms existing designs adopting the i.i.d. channel assumption.
	
%	\item The maximum KGR under the assumption of spatial correlated model is higher than that under i.i.d. assumption.   
	%	\item The impact of spatial correlation on the secret key rate is investigated. It is found that the correlation at both BS and RIS has a positive impact on the maximum key generation rate (KGR) in our proposed framework. In addition, with the number of antennas growing large, the maximum KGR converages.
	\item Simulation results show that compared with existing
	methods that ignore the spatial correlation, the optimal design achieves about $5$ dB gain when the antenna correlation coefficient is $0.3$ and the element spacing is half of the wavelength.
	Moreover, the KGR gain increases with the spatial correlation at both the BS and the RIS.
\end{itemize}

\section{System Model}\label{Sec-Model}
\begin{figure}
	\centering
	\includegraphics[width=3.1in]{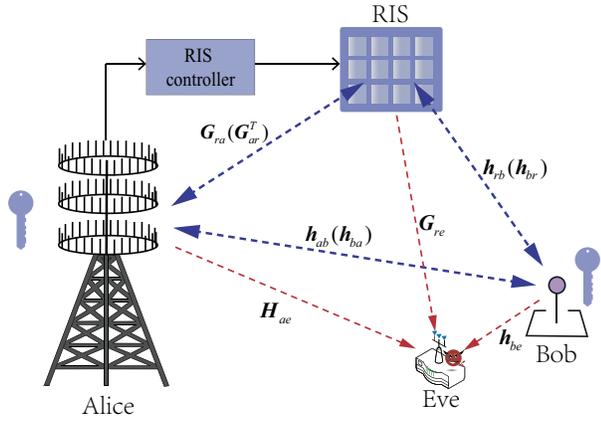}
	\caption{The model of RIS-assisted PKG based on transmit and reflective beamforming.}
	\label{system model}
\end{figure}
%As shown in Fig. \ref{system model}, we investigates a RIS-assisted point-to-point key generation in a multi-antenna system, where the spatial correlation channel models at BS and RIS are both taken into account. 
%In particular, we consider a MISO a multi-antenna access point Alice and a single-antenna Bob aim to generate symmetric key from the channel between them,   while eavesdropper Eve intends to obtain the key information form her received signals. 
As shown in Fig. \ref{system model}, we study a RIS-assisted key generation method in a multiple-input single-output multi-antenna eavesdropper (MISOME) system, in which a multi-antenna base station (BS), Alice, and a single-antenna user, Bob, aim to generate symmetric keys \cite{Ji},\cite{SPletter} from the wireless channel with the help of a RIS adopting the time-division duplexing (TDD) protocol. Meanwhile, a multi-antenna eavesdropper, Eve, intends to obtain the key information from her received signals. 
%Specifically, assuming a time-division duplexing (TDD) mode, Alice and Bob sound channel alternatively in coherence time to obtain the reciprocal channel estimation.
We assume that Alice and Eve are equipped with $ M $ and $ K $ antennas, respectively. The RIS consists of $N$ passive reflecting elements and introduces phase shifts to the impinging signals to facilitate key generation. 
Since the spatial correlation affects the secret key rate, we consider the general spatial correlation channel model at both the RIS and the BS. 
%	which
%	cannot be ignored especially for limited size of RIS
%	To describe channel characteristics more accurately, 

%Alice, Bob and Eve are equipped with $M$ antennas, single antenna and $J$ antennas, respectively. 

%In this paper, we assume that the direct channel batween Alice and Bob is blocked and a RIS consistig of $N$ passive elements is deployed to establish a virtual LoS link to enable PKG. The RIS reflection coefficients are designed at BS and sent to RIS controller through the backhaul link. 
%The key generation 

\subsection{Channel Model}
%the spatial correlation of RIS is neglected
%Let $\boldsymbol{G}_{ra}\in \mathbb{C}^{M \times N}$, $\boldsymbol{h}_{b r}\in 
%\mathbb{C}^{N \times 1}$, and $\boldsymbol{H}_{e r} = [ \boldsymbol{h}_{1r}, \cdots, \boldsymbol{h}_{Kr}, ]\in \mathbb{C}^{N \times K}$ be the channel from RIS to Alice, from Bob to RIS, and from Eve to RIS, respectively. 

The direct channels of Alice-to-Bob, Eve-to-Bob, and Alice-to-Eve are denoted by $\boldsymbol{h}_{ab}  \in \mathbb{C}^{M \times 1}$, 
$\boldsymbol{h}_{eb}\in \mathbb{C}^{K \times 1}$, and $\boldsymbol{H}_{ae}  \in \mathbb{C}^{K \times M}$, respecively, where $\mathbb{C}^{A\times B}$ denotes the space of complex matrices of size $A\times B$. $\boldsymbol{h}_{ak} \in \mathbb{C}^{M \times 1}$ denotes the channel from Alice to Eve's $k$-th antenna, $k\in \{1,\cdots,K\}$. 
When a RIS is involved in the PKG system, it introduces additional channels.
Specifically, the channels of RIS-to-Alice, RIS-to-Bob, and RIS-to-Eve are represented as $\boldsymbol{G}_{ra}\in \mathbb{C}^{M \times N}$, $\boldsymbol{h}_{r b}\in \mathbb{C}^{N \times 1}$, and $\boldsymbol{G}_{r e} \in \mathbb{C}^{K \times N}$, respectively.  $\boldsymbol{h}_{r k} \in \mathbb{C}^{M \times 1}$ denotes the channel from the RIS to Eve's $k$-th antenna. 
%where $ \boldsymbol{h}_{kr} $ for $k = 1, \cdots, K$ in $ \boldsymbol{H}_{e r} $ is the channel from Eve's $k$-th antenna to RIS. 
%The direct channels $ \boldsymbol{h}_{ba} \in \mathbb{C}^{M \times 1} $, $ \boldsymbol{h}_{be}=[h_{b1}, \cdots,h_{bK}]^{T} \in \mathbb{C}^{K \times 1} $, and $ \boldsymbol{H}_{ea}=[\boldsymbol{h}_{1,a},\cdots,\boldsymbol{h}_{K,a} ] \in \mathbb{C}^{M \times K} $ denote the channel from Bob to Alice, from Bob to Eve, and from Eve to Alice, respectively. 
To account for the spatial correlation, the channel matrices are described by employing the Kronecker correlation channel model as 
\begin{align}
	\boldsymbol{G}_{ra} &=\boldsymbol{G}_{ar}^T= \beta_{ar}^{\frac{1}{2}}
	\boldsymbol{R}_{S}^{\frac{1}{2}} \tilde{ \boldsymbol{H}} \boldsymbol{R}_{I}^{\frac{1}{2}}
	, \label{eq1}\\
	\boldsymbol{h}_{ri}&=\boldsymbol{h}_{ir} =  \beta_{ir}^{\frac{1}{2}}\boldsymbol{R}_{I}^{\frac{1}{2}} \tilde{ \boldsymbol{h}}_{ir}, \ i \in \{b, k\}, \label{eq2} \\
	\boldsymbol{h}_{aj}&=\boldsymbol{h}_{ja}= \beta_{ja}^{\frac{1}{2}}\boldsymbol{R}_{S}^{\frac{1}{2}} \tilde{ \boldsymbol{h}}_{ja}, \ j \in \{b,k\}, \label{eq3}
\end{align}
respectively, where $\boldsymbol{R}_{S} \in \mathbb{C}^{M \times M}$ and $\boldsymbol{R}_{I} \in \mathbb{R}^{ N \times N} $ are the spatial correlation matrices at Alice and the RIS, respectively\cite{Rayleigh-fading}.
% $\tr(\cdot)$ denotes the trace of a matrix. 
The elements $\left[ \boldsymbol{R}_{S}\right]_{m,n}$ and $\left[ \boldsymbol{R}_{I}\right]_{m,n}$ represent the correlation between the $m$-th antenna/element and the $n$-th antenna/element. In addition, 
$\tilde{ \boldsymbol{H}} \in \mathbb{C}^{M \times N}$, $\tilde{ \boldsymbol{h}}_{ir} \in \mathbb{C}^{N \times 1}$, and $\tilde{ \boldsymbol{h}}_{ja} \in \mathbb{C}^{M \times 1}$ are random matrices with i.i.d. Gaussian random entries of zero mean and unit variance. 
$ \beta_{ar}$, $\beta_{ir}$, and $\beta_{ja}$ are the path loss of the corresponding channels, respectively.  

%The elements in the normalized spatial correlation matrix $ \boldsymbol{R}_{I} $ is 
%\begin{align}
%	[\mathbf{R}]_{n, m}=\operatorname{sinc} \frac{2\left\|\mathbf{u}_{n}-\mathbf{u}_{m}\right\|}{\lambda} \quad n, m=1, \ldots, N
%\end{align}
%where the sinc function is expressed as $ \operatorname{sinc}(t)=\operatorname{sin}(\pi t)/(\pi t) $. 
%
%
%\cite{Rayleigh-fading}

\subsection{PKG Framework Based on Transmit and Reflective Beamforming} 
%In PKG system, Alice and Bob treat the reciprocal wireless channel as common randonmess to extract a symmetric key. 
%The general key generation process consists four steps, i.e. , while in this paper we focus on the first step to maximize the theoretical key rate. 
Now, we propose a new framework to take full advantages of the RIS-assisted PKG in multi-antenna systems. 
%In this paper, 
%Take full advantage of the RIS-assisted PKG in multi-antenna systems,
%%Different from the SISO system that only the RIS coefficients are configured to improve the key generation rate, 
%we propose to joint design the transmit and reflective beamforming at BS and RIS, respectively. 
%The general PKG contains four steps, namely channel probing, quantization, information reconciliation, and privacy amplification. Since the last three steps are similar to existing works [access], in this paper we focus on the channel probing step where transmit and reflective beamforming are joint designed to facilitate PKG. 
In PKG, Alice and Bob first perform channel probing to acquire the reciprocal channel estimation. 
The process of channel probing is described as follows.

\emph{Step 1: Uplink channel sounding.} Bob transmits the publicly known pilot $s_{u} \in\mathbb{C} $ with $s_{u}^{*}s_{u}=1$. Then, the equivalent baseband signal received at Alice and Eve are expressed as 
\begin{align}
	\boldsymbol{y}_{l}^{u} =  \sqrt{P_\mathrm{B}} \left(\boldsymbol{G}_{rl}  \boldsymbol{\Phi}\boldsymbol{h}_{br} + \boldsymbol{h}_{bl}    \right) s_{u} + \boldsymbol{z}_{l}, l\in \{a,e\},
\end{align}
respectively, where $\boldsymbol{\Phi} = \diag\{\boldsymbol{v}\}$ with each element $|v_{n}|=1, \forall n\in  \{1,\cdots,N\}$, is the reflection cofficients matrix of the RIS. 
$\diag(\boldsymbol{x})$ and
$|x|$ denotes a diagonal matrix 
and the modulus of a complex scalar, respectively. In addition, $P_\mathrm{B}$ is the transmit power of Bob. The noise follows the circularly symmetric complex Gaussian distribution, i.e., 
 $\boldsymbol{z}_{a} \sim \mathcal{CN}(0, \sigma_{a}^2 \boldsymbol{I}_{M \times M})$, $\boldsymbol{z}_{e} \sim \mathcal{CN}(0, \sigma_{e}^2 \boldsymbol{I}_{K \times K})$, where $\sigma_{ a }^2$ and $\sigma_{ e }^2$ are
the noise variances of Alice and Eve, respectively. Then, Alice and Eve perform the least square (LS) estimation\footnote{The LS is adopted since it has been widely used in practical systems\cite{li2021maximizing}.} as
\begin{align}
	\hat{\boldsymbol{h}}_{l}^{u} \triangleq  s_{u}^* \boldsymbol{y}_{l}^{u} 
	= \sqrt{P_\mathrm{B}} (\boldsymbol{G}_{rl} \boldsymbol{\Phi}\boldsymbol{h}_{br} +   \boldsymbol{h}_{bl}) + \tilde{\boldsymbol{z}}_{l}^{u}, l\in \{a,e\}, \label{ha1} 
\end{align} 
respectively, where the estimation noise is $ \tilde{\boldsymbol{z}}_{l}^{u}=s_{u}^*\boldsymbol{z}_{l}^{ u }  $.

\emph{Step 2: Downlink channel sounding.} Alice sends the pilot $s_{d} \in \mathbb{C}$ with $s_{d}^*s_{d} = 1 $, and the signals received at Bob and Eve are  
\begin{align}
	y_{b}^{d} &=  (\boldsymbol{h}_{rb}^T \boldsymbol{\Phi} \boldsymbol{G}_{ar}  +  \boldsymbol{h}_{ab}^{ T } )\boldsymbol{w} s_{d} +  z_{b}^{d}, \\
	\boldsymbol{y}_{e}^{d} &=  (\boldsymbol{G}_{re} \boldsymbol{\Phi} \boldsymbol{G}_{ar}  +  \boldsymbol{H}_{ae} )\boldsymbol{w} s_{d} +  \boldsymbol{z}_{e}^{d}, 
\end{align}
respectively, where $\boldsymbol{w}$ is the transmit beamforming vector at Alice that satisfies $||\boldsymbol{w}||^2 \leq P_\mathrm{A}$. $||\cdot||$ denotes the Euclidian norm. $z_{b}^{d}$ and $\boldsymbol{z}_{e}^{d} $ are the additive Gaussian noise with $z_{b}^{d} \sim \mathcal{CN}\left(0, \sigma_{b}^2\right)$ and $\boldsymbol{z}_{e}^{d} \sim \mathcal{CN}\left(0, \sigma_{e}^2\boldsymbol{I}_{K \times K}\right)$. 
After the LS estimation, Bob and Eve obtain the channel estimates as 
\begin{align}
	\hat{ h}_{ b }\triangleq s_{d}^* y_{b}^{d} 
	&= (\boldsymbol{h}_{rb}^{T}\boldsymbol{\Phi} \boldsymbol{G}_{ar}  +  \boldsymbol{h}_{ab}^{ T } )\boldsymbol{w} + \tilde{z}_{b},  \label{hb} \\
	\hat{\boldsymbol{h}}_{ e }^{d} \triangleq s_{d}^* \boldsymbol{y}_{e}^{d} 
	&= (\boldsymbol{G}_{re} \boldsymbol{\Phi} \boldsymbol{G}_{ar}  +  \boldsymbol{H}_{ae} )\boldsymbol{w} + \tilde{\boldsymbol{z}}_{e}^{u},
\end{align}
respectively, where the noises are $\tilde{z}_{b}^{ d } = s_{d}^* z_{b}^{d} $ and $\tilde{\boldsymbol{z}}_{e}^{u}=s_{d}^*\boldsymbol{z}_{e}^{u}$, respectively.

\emph{Step 3: Reciprocal components acquisition.} 
Since the estimations obtained by Alice and Bob, as shown in (\ref{ha1}) and (\ref{hb}), are quite different, we multiply Alice's channel estimation $ \hat{\boldsymbol{h}}_{a}^{u}$ by $\boldsymbol{w}$ to obtain the combined reciprocal channel gain as
\begin{align}
	\hat{ h}_{a}  \triangleq  \boldsymbol{w}^T  \hat{\boldsymbol{h}}_{a}^{u}= \sqrt{P_\mathrm{B}} \boldsymbol{w}^T ( \boldsymbol{G}_{ra} \boldsymbol{\Phi}\boldsymbol{h}_{br}+   \boldsymbol{h}_{ba} )+ z_{a}, \label{ha}
\end{align}
where the noise is $z_{a}=\boldsymbol{w}^T \tilde{\boldsymbol{z}}_{a}^{u}$. 

Consequently, Alice's combined channel gain, $\hat{ h}_{a}$, and Bob's channel gains, $\hat{ h}_{b}$, are highly correlated. 
After the following procedures of the PKG, i.e., quantization, information reconcilation, and privacy amplification, the channel gains are finally converted into secret keys \cite{li2021reconfigurable}. Since
these steps are similar to those used in existing PKG methods, in this paper, we focus on 
the channel probing step, where the transmit and reflective beamforming are optimized to maximize the KGR.

\section{Problem Formulation}
%In Ji TvT, multiple eve are assumed in the system and all of their channels are correlated with the legitimate channel. But in fact, only when the Eve locates Bob less than half wavelength, its channel is correlated. Therefore, in this papr we assume a Evadropper is close to Bob and is equipped with muitiple antennas. 
In this section, we formulate an optimization problem to find the optimal transmit beamforming $\boldsymbol{w}$ and the reflective beamforming $\boldsymbol{v}$ by deriving the closed-form KGR expression. 

%we derive a closed-form expression for the secret key rate based on the proposed framework in Sec. \ref{Sec-Model}-B. 
First, the secret key rate is defined as the conditional mutual information of legitimate parties' channel estimations given the observation of Eve \cite{li2021maximizing}, which is expressed as 
\begin{align}
%	R_{SK}  \triangleq \mathcal{I}\left(\tilde{ h}_{a} ; \tilde{ h}_{b}|  \overline{h}_{e1},\cdots, \overline{h}_{eK},  \tilde{ h}_{ e1 },\cdots, \tilde{ h}_{ eK } \right),
R_{\mathrm{SK}}  \triangleq \mathcal{I}\left(\hat{ h}_{a} ; \hat{ h}_{b}|  \hat{\boldsymbol{h}}_{e}^{u}, \hat{\boldsymbol{h}}_{e}^{d} \right),
\end{align}
where $\mathcal{I}(X;Y)$ is the mutual information of random variables $X$ and $Y$. 
In this paper, we assume that Eve is located at least half-wavelength away from Alice and Bob. 
%Hence, the channels from legitimate parties to Eve are independent of the channel between Alice and Bob
Hence, the eavesdropping channels are independent of the legitimate channels\footnote{Due to the space limitation, the case where Eve experiences a correlated channel will be investigated in the extended journal version.}. In this case, the KGR is given by \cite{li2021maximizing}
\begin{align}
	R_{\mathrm{SK}} = \mathcal{I}\left(\hat{ h}_{a} ; \hat{ h}_{b} \right) = \log_{2} \frac{\mathcal{R}_{a,a} \mathcal{R}_{b,b}}{\operatorname{det}\left(\boldsymbol{R}_{ab}\right)}, \label{R}
\end{align}
where $\mathcal{R}_{i,j}=\mathbb{E}\{\hat{ h}_{i}\hat{ h}_{j}^H\},i,j\in \{a,b\}$, $\operatorname{det}(\cdot)$ is the matrix determinant, $\mathbb{E}\{ \cdot \}$ denotes the statistical expectation, and
\begin{align}
	\boldsymbol{R}_{ab} =\left[\begin{array}{cc}
		\mathcal{R}_{a,a} & \mathcal{R}_{a,b} \\
		\mathcal{R}_{b,a} & \mathcal{R}_{b,b}
	\end{array}\right].
\end{align}
%
% $\mathcal{R}_{b}$, and $ \mathcal{R}_{a b} $ are  
%\begin{align}
%	\mathcal{R}_{p_{1}, ,\cdots, p_{L}  }  \triangleq \mathbb{E}\left\{  \boldsymbol{p} \boldsymbol{p}^H      \right\},\  \boldsymbol{p} = [p_{1}, p_{2}, \cdots, p_{L} ]^T. 
%\end{align}
Substituting the channel estimations into (\ref{R}) and assuming $\sigma_{ a }^2 = \sigma_{ b }^2=\sigma^2$ for simplicity, $ R_{\mathrm{SK}} $ is expressed as (\ref{eq:whole}) at the top of the next page, where $ \tilde{\boldsymbol{R}}_{I} = \boldsymbol{R}_{I} ^{ T } \circ   \boldsymbol{R}_{I}  $, $\beta_{ r }=\beta_{ar}\beta_{br}$, and $\circ$ denotes Hadamard product.
%\begin{align}
%	 R_{\mathrm{SK}} = \log_{2}
%	\left(f_{0}(a)        \right) + \log_{2}
%	\left(f_{1}( b  )        \right) 
% 	- \log_{2}
%	\left(f_{2}( a,b  )        \right),
%\end{align}
%where
%\begin{align}
%	f_{0}(a ) & =  
%	P_B\boldsymbol{ w }^{ T } \boldsymbol{R}_S \boldsymbol{w}^*(\beta_{ r }  \boldsymbol{v}^H \tilde{\boldsymbol{R}}_{I}\boldsymbol{v} + \beta_{ ba } ) +  ||\boldsymbol{w}||^2 \sigma_{ a }^{ 2 },
%	 \\%
%	f_{1}(b ) & =  \boldsymbol{ w }^{ T } \boldsymbol{R}_S \boldsymbol{w}^*(\beta_{ r }  \boldsymbol{v}^H \tilde{\boldsymbol{R}}_{I}\boldsymbol{v} + \beta_{ ba } ) +  \sigma_{ b }^{ 2 }, \\%
%	f_{2}(a,b ) &=  
%	(P_A\sigma_{ a }^{ 2 } + P_B\sigma_{ b }^2)\boldsymbol{ w }^{ T } \boldsymbol{R}_S \boldsymbol{w}^*(\beta_{ r }  \boldsymbol{v}^H \tilde{\boldsymbol{R}}_{I}\boldsymbol{v} + \beta_{ ba } )  \notag\\ 
%	& \quad + P_A\sigma_{ a }^2 \sigma_{ b }^2,
%\end{align}

\begin{IEEEproof}
	See Appendix~\ref{sec:proof_of_thm1}. 
\end{IEEEproof}

%\begin{figure*}[hb]
%	\centering
%	\begin{align}
%		\rule{\textwidth}{0.1mm} \notag \\
%		I_{j} = \log_{2} \frac{\kappa_{a}^2\kappa_{b}^2}{\sigma_{a}^2\kappa_{a}^2 + \sigma_{b}^2\kappa_{b}^2 }
%		\left( R_{j} - \frac{\sigma_{b}^2}{\kappa_{a}^2} + \frac{\kappa_{a}^2\sigma_{a}^4}{\kappa_{b}^2(\sigma_{a}\kappa_{a}^2 + \sigma_{b}^2\kappa_{b}^2)} + \frac{\sigma_{a}^4\sigma_{b}^4}{(\kappa_{a}^2\sigma_{a}^2+\kappa_{b}^2\sigma_{b}^2)^2R_{j}+(\kappa_{a}^2\sigma_{a}^2+\kappa_{b}^2\sigma_{b}^2)\sigma_{a}^2\sigma_{b}^2 }  
%		\right)
%	\end{align}
%\end{figure*}

%%------------跨栏公式
\newcounter{mytempeqncnt}
\begin{figure*}[!t]
	\normalsize
	% \setcounter{mytempeqncnt}{\value{equation}}
	% \setcounter{equation}{10}
%	\begin{equation}\label{eq:whole}
%		\mathcal{I}\left(\tilde{ h}_{a} ; \tilde{ h}_{b}\right)=
%		\log \frac{ \left(\beta_{ r } \boldsymbol{ w }^{ T } \boldsymbol{R}_s \boldsymbol{w}^* \boldsymbol{v}^H \tilde{\boldsymbol{R}}_{I}\boldsymbol{v} + \beta_{ d } \boldsymbol{ w }^{ T } \boldsymbol{R}_s \boldsymbol{w}^* +  \sigma^{ 2 }\right)^2}
%{   2\sigma^2 \beta_{ r } \boldsymbol{ w }^{ T } \boldsymbol{R}_s \boldsymbol{w}^* \boldsymbol{v}^H \tilde{\boldsymbol{R}}_{I}\boldsymbol{v} + 2\sigma^2\beta_{ d } \boldsymbol{ w }^{ T } \boldsymbol{R}_s \boldsymbol{w}^* + \sigma^4        }
%	\end{equation}
\begin{equation}\label{eq:whole}
	R_{\mathrm{SK}}= 
	\log_{2} \frac{ 
		(P_\mathrm{B}\boldsymbol{ w }^{ T } \boldsymbol{R}_S \boldsymbol{w}^*(\beta_{ r }  \boldsymbol{v}^H \tilde{\boldsymbol{R}}_{I}\boldsymbol{v} + \beta_{ ba } ) +  ||\boldsymbol{w}||^2 \sigma^{ 2 })
	(\boldsymbol{ w }^{ T } \boldsymbol{R}_S \boldsymbol{w}^*(\beta_{ r }  \boldsymbol{v}^H \tilde{\boldsymbol{R}}_{I}\boldsymbol{v} + \beta_{ ba } ) +  \sigma^{ 2 })  }
	{   (||\boldsymbol{w}||^2+ P_\mathrm{B})\sigma^2\boldsymbol{ w }^{ T } \boldsymbol{R}_S \boldsymbol{w}^*(\beta_{ r }  \boldsymbol{v}^H \tilde{\boldsymbol{R}}_{I}\boldsymbol{v} + \beta_{ ba } )   + ||\boldsymbol{w}||^2 \sigma^4      }.
\end{equation}
	\hrulefill
\end{figure*}
%As observed in the secret key rate expression, the rate is only dependent on the long-term channel state information, i.e. the path loss $ \beta_{d} $, $ \beta_{ r } $ and the covariance matrices $ \boldsymbol{R}_{s} $ and $ \boldsymbol{R}_{I} $. 
Thus, the beamforming design can be formulated as 
\begin{align}
	\mathcal{P}:\underset{\boldsymbol{w}, \boldsymbol{v}}{\operatorname{maximize}} \   \ &R_{\mathrm{SK}}
	\label{P0} \\
	\operatorname{subject\ to}\ &|v_{n} |= 1 , \forall n\in \{1,\cdots N\}, \tag{\ref{P0}{a}} \label{P0-a} \\
	& ||\boldsymbol{w}||^2 \leq P_\mathrm{A}, \tag{\ref{P0}{b}} \label{P0-b} 
\end{align}
where (\ref{P0-a}) 
	represents the unit modulus constraint of each reflection coefficient, while  (\ref{P0-b}) indicates the transmit beamforming is constrained by the maximum transmit power.

%% 算法设计部分
\section{Proposed Solution to Problem $\mathcal{P}$}
In this section, we jointly optimize the transmit beamforming $ \boldsymbol{w} $ and the reflective beamforming $ \boldsymbol{v} $ to maximize the KGR. 
%Specifically, by exploiting the structure of $ \mathcal{P} $, we could find the closed-form global optimal solution. 
To tackle the non-convex problem in (\ref{P0}), we decompose the problem into two sub-problems 
%in terms of $\boldsymbol{w}$ and $\boldsymbol{v}$, 
and optimize them to obtain the globally optimal solution.

\subsection{Problem Decomposition}
It could be found that in problem $ \mathcal{P} $, the objective function (\ref{eq:whole}) contains high-order terms in $ \boldsymbol{w} $ and $ \boldsymbol{v} $, while the unit modulus constraint (\ref{P0-a}) and quadratic equality constraints (\ref{P0-b}) are both non-convex and the optimization variables are coupled.
To tackle this problem, we first decompose the problem into two sub-problems with respect to $ \boldsymbol{w} $ and $ \boldsymbol{v}$,  respectively, using the following Lemma. 
\begin{lemma} \label{lemma1}
	The objective function in $ \mathcal{P} $ increases monotonically with $ \boldsymbol{ w }^{ T } \boldsymbol{R}_s \boldsymbol{w}^*(\beta_{ r }  \boldsymbol{v}^H \tilde{\boldsymbol{R}}_{I}\boldsymbol{v} + \beta_{ ba } ) $. 
\end{lemma}
\begin{IEEEproof}
	See Appendix B.
\end{IEEEproof}
%Therefore, the problem $\mathcal{P}$ could be globally optimized by maximizing these two terms with respect to $\boldsymbol{w} $ and $ \boldsymbol{v}  $, respectively. 
Since $\boldsymbol{ w }^{ T } \boldsymbol{R}_S \boldsymbol{w}^*$ and $(\beta_{ r }  \boldsymbol{v}^H \tilde{\boldsymbol{R}}_{I}\boldsymbol{v} + \beta_{ ba } )$ are both positive, solving problem $\mathcal{P}$ is equivalent to maximize these two terms separately.

\subsection{Transmit Beamforming Optimization}

%Problem $\mathcal{P}$ is equivalent to 
%\begin{align}
%	\mathcal{P}:\underset{\boldsymbol{w}, \boldsymbol{v}}{\operatorname{maximize}} \   \ & \boldsymbol{ w }^{ T } \boldsymbol{R}_s \boldsymbol{w}^*(\beta_{ r }  \boldsymbol{v}^H \tilde{\boldsymbol{R}}_{I}\boldsymbol{v} + \beta_{ d } )
%	\label{P1} \\
%	\operatorname{ subject\ to }\ &|v_{n} |= 1 , \forall n= 1, 2, \cdots, N , \tag{\ref{P1}{a}} \label{P1-a} \\
%	& ||\boldsymbol{w}||^2 = P_A, \tag{\ref{P1}{b}} \label{P1-b}  
%\end{align}
%%\begin{lemma}\label{lemma1}
%\begin{IEEEproof}
%	See Appendix~\ref{sec:proof_B}.
%\end{IEEEproof}
%Since the $ \beta_{ r }  \boldsymbol{v}^H \tilde{\boldsymbol{R}}_{I}\boldsymbol{v} + \beta_{ d } >0 $ for $ \forall \boldsymbol{v} $. 
%From lemma 1, the optimization problem to obtain the optimal transmit beamforming is expressed as
One of the sub-problems is to optimize the transmit beamforming vector, which is expressed as $\mathcal{P}_{1}$ by denoting $\bar{\boldsymbol{w}} = \boldsymbol{w}^*$:
\begin{align}
	\mathcal{P}_1: \underset{ \bar{\boldsymbol{w}}}{\operatorname{maximize}} \   \ & 
	\bar{\boldsymbol{w}}^H \boldsymbol{R}_S \bar{\boldsymbol{w}} \label{P1}  \\
	\operatorname{subject\ to}\ & ||\bar{\boldsymbol{w}}||^2 \leq P_\mathrm{A} \tag{\ref{P1}{a}}. \label{P1-a} 
\end{align}
%\end{lemma}
By using the Rayleigh quotient, the optimal solution to $\mathcal{P}_1$ is
\begin{align}
	\bar{\boldsymbol{w}}_{ \mathrm{opt} } = \sqrt{P_\mathrm{A}} \boldsymbol{u}_{\lambda_{\mathrm{max}}}, \label{wopt}
\end{align}
where $\boldsymbol{u}_{\lambda_{\mathrm{max}}}$ is the 
dominant eigenvector of the matrix $\boldsymbol{R}_S$ corresponding to its maximum eigenvalue $\lambda_{\mathrm{max}}$. 
%In this case, the optimal value of $ \mathcal{P}_1$ is $ \boldsymbol{w}^T \boldsymbol{R}_s \boldsymbol{w}^* = P_{A} \lambda_{m} $. 

\subsection{Reflective Beamforming Optimization}
After deriving the optimal transmit beamforming vector, we aim to optimize the reflection coefficients at RIS. With Lemma 1, the problem of optimizing $ \boldsymbol{v} $ is equivalent to
\begin{align}
	\mathcal{P}_2: \underset{ \boldsymbol{v}}{\operatorname{maximize}} \   \ & 
	\boldsymbol{v}^H  \tilde{\boldsymbol{R}}_{I} \boldsymbol{v}  \label{P2} \\
	\operatorname{subject\ to}\ & |v_{n} |= 1 , \forall n\in \{1,\cdots, N\}.  \tag{\ref{P2}{a}} \label{P2-a} 
\end{align} 
%It is noted that $ \mathcal{P}_2 $ is to maximize a convex quadratic term. Moreover, 
It is noted that the unit modulus constraints in (\ref{P2-a}) are intrinsically non-convex \cite{li2021maximizing}. Therefore, it is challenging to solve this problem.
%	This non-convex constraint together with the non-convex objective function makes  difficult to solve. 
%Therefore, problem $ \mathcal{P}_2 $ is NP-hard \cite{NPhard}. 
Nevertheless, we note that each element in $ \tilde{\boldsymbol{R}}_{I} = \boldsymbol{R}_{I} ^{ T } \circ   \boldsymbol{R}_{I} $ is a positive number since the covariance matrix $\boldsymbol{R}_{I}$ is real symmetric. 
Based on this observation, the optimal solution is given as follows. 
\begin{theorem}
	The optimal solution to problem $ 	\mathcal{P}_2 $ is the case where all elements of $ \boldsymbol{v} $ adopt the same phase, 
	i.e.,
	\begin{align}
		\theta_{n} = \theta, \ \forall n \in \{1, \cdots, N\},
	\end{align}
where $ \theta $ could take on any value in interval $[0, 2\pi)$. 
\end{theorem}

\begin{IEEEproof}
	The objective function (\ref{P2}) could be calculated as 
	\begin{align}
		\boldsymbol{v}^H  \tilde{\boldsymbol{R}}_{I} \boldsymbol{v} &= \sum_{n = 1}^{N }[\tilde{\boldsymbol{R}}_{I}]_{n, n} 
		+ \sum_{j = 1}^{N} \sum_{i = 1}^{ N }[\tilde{\boldsymbol{R}}_{I}]_{i,j}  v_{i} v_{j}^*    \notag \\
		& = \sum_{n = 1}^{N }[\tilde{\boldsymbol{R}}_{I}]_{n, n} 
		+ \sum_{j = 1}^{N} \sum_{i = 1,i > j}^{ N }2 [\tilde{\boldsymbol{R}}_{I}]_{i,j}  \cos{ (\theta_{i} - \theta_{j}) }.  
	\end{align} 
%It can be found that when all reflecting elements are set to the same reflection phase, the objective function obtains maximum value. 
Since $\cos{(\theta_{i} - \theta_{j})}\leq 1$, the maximum value could be obtained when $\theta_{i} = \theta_{j},\ \forall i,j$.
This completes the proof.  
%The optimal value is 
%\begin{align}
%	\boldsymbol{v}^H_{opt}  \tilde{\boldsymbol{R}}_{I} \boldsymbol{v}_{opt} = \sum_{j = 1}^{ N } \sum_{i = 1}^{N} [\tilde{\boldsymbol{R}}_{I}]_{i,j} = ||\boldsymbol{R}_{I} ||_{F}^{2} 
%\end{align}
\end{IEEEproof}
In the case of optimal $\boldsymbol{w}$ and $\boldsymbol{v}$, the KGR only depends on the large-scale path loss, indicating the maximum KGR is dependent on the distance between Alice, Bob, and the RIS. Moreover, in the spatially correlated channel model, the optimal beamforming are determined by the spatial correlation matrices at the BS and the RIS, which could be obtained effectively by existing methods, such as \cite{Rayleigh-fading} and \cite{neumann2018covariance}.

\section{Impact of Different Beamforming Methods on PKG Performance}
%In Sec. II, we have obtained the optimal solution for reaching the maximum KGR, which is determined by the spatial correlation matrices. 
In this section, we aim to compare the PKG performance under the assumptions of the i.i.d. channel model and the spatially correlated channel model.  

\subsection{KGR under Different Assumptions of Channel Model at BS}
As shown in Lemma 1, the KGR is proporational to $\boldsymbol{w}^T \boldsymbol{R}_{S} \boldsymbol{w}^*$.  
Under the assumption of the i.i.d. model, the spatial correlation matrix $\boldsymbol{R}_{S}$ is considered as an identity matrix. In this case, the design of transmit beamforming is independent of KGR. As such, random beamforming $\widetilde{\boldsymbol{w}}=\sqrt{P_\mathrm{A}}\widetilde{\boldsymbol{w}}_{0}/||\widetilde{\boldsymbol{w}}_{0}|| $ is applied without loss of generality, where the entries in  $\widetilde{\boldsymbol{w}}_{0}$ are i.i.d. random variables with zero mean. Then, the expectation of the objective function in $\mathcal{P}_{1}$ is calculated as $\mathbb{E}\{\widetilde{\boldsymbol{w}}^T \boldsymbol{R}_{S} \widetilde{\boldsymbol{w}}^*\} = P_\mathrm{A}$, 
%	\begin{align}
%		\mathbb{E}\{\widetilde{\boldsymbol{w}}^T \boldsymbol{R}_{S} \widetilde{\boldsymbol{w}}^*\} = P_{A},
%	\end{align}
which is independent of the antenna number and the spatial correlation at the BS. 

To investigate the performance loss caused by the design based on the i.i.d. channel assumption, we focus on a typical implementation model of multiple antennas for massive multiple-input multiple-output (MIMO). We consider a general uniform planar array (UPA) model, where the spatial correlation matrix can be approximated as $\boldsymbol{R}_{S} \approx \boldsymbol{R}_{h} \otimes \boldsymbol{R}_{v}$ \cite{choi2014bounds},
%\begin{align}
%	\boldsymbol{R}_{S} \approx \boldsymbol{R}_{h} \otimes \boldsymbol{R}_{v},  \label{RU}
%\end{align} 
 where $ \boldsymbol{R}_{h} $ and $ \boldsymbol{R}_{v} $ are the covariance matrices of the horizontal and the vertical uniform linear array (ULA), respectively. 
%Then, by adopting the exponential correlation model that $ [\boldsymbol{R}_{l}]_{n,m} = \rho^{| n - m|} $, we have the following lemma.
%For the ULA spatial correlation matrix, we could model the $ \boldsymbol{R}_{l} $ as an exponential model as a Toeplitz matrix with each element $ \rho^{| i-j|} $, which is the common way to model the ULA spatial correlation matrices. 
The ULA spatial correlation is modeled as a Toeplitz matrix with each element $[\boldsymbol{R}_{l}]_{i,j} = \rho^{| i-j|},l\in \{h,v\} $, where $ \rho $ is the correlation index among the antennas. 
Given the optimal transmit beamforming (\ref{wopt}), we have the following lemma. 
\begin{lemma}
	For a UPA model, the upper and lower bounds for the $\boldsymbol{w}_{\mathrm{opt}}^T \boldsymbol{R}_{S} \boldsymbol{w}_{\mathrm{opt}}^*$ are given by 
	\begin{align}
		 f_{l}(N_{\mathrm{H}}^{t},N_{\mathrm{V}}^{t},\rho)  \leq  \boldsymbol{w}_{\mathrm{opt}}^T \boldsymbol{R}_{S} \boldsymbol{w}_{\mathrm{opt}}^* \leq   f_{u}(N_{\mathrm{H}}^{t},N_{\mathrm{V}}^{t},\rho), 
	\end{align}
	where $ f_{l}(N_{\mathrm{H}}^{t},{N_{\mathrm{V}}^{t}},\rho) = P_\mathrm{A}\frac{ \left( {N_{\mathrm{H}}^{t}}(1 - \rho^2) - 2\rho(1 - \rho^{N_{\mathrm{H}}^{t}})   \right)  }{ {N_{\mathrm{H}}^{t}}{N_{\mathrm{V}}^{t}}(1 - \rho)^4 } \times \left( {N_{\mathrm{V}}^{t}}(1 - \rho^2) - 2\rho(1 - \rho^{N_{\mathrm{V}}^{t}})   \right)$ and $ f_{u}({N_{\mathrm{H}}^{t}},{N_{\mathrm{V}}^{t}},\rho) =P_\mathrm{A} \frac{ (1+\rho^2)(1 - \rho^{{N_{\mathrm{H}}^{t}}-1} )(1 - \rho^{{N_{\mathrm{V}}^{t}} - 1} ) }{ (1 - \rho)^2 } $. $ N_{\mathrm{H}}^{t}$ and $N_{\mathrm{V}}^{t}$ are the number of antennas at horizontal and vertical domains, respectively.
\end{lemma}
\begin{IEEEproof}
	See Appendix~\ref{section:proof-lemma3} 
\end{IEEEproof}
This lemma shows that the both the upper and lower bounds increase monotonically with the correlation coefficients $\rho$, the number of antennas $N_{\mathrm{H}}^{t}$, and $N_{\mathrm{V}}^{t}$. This is because the SNR of the combined channel gain increases with the spatial correlation. 
Specifically, when $\rho=0$, the bounds are  $f_{l}({N_{\mathrm{H}}^{t}},{N_{\mathrm{V}}^{t}},0)=f_{u}({N_{\mathrm{H}}^{t}},{N_{\mathrm{V}}^{t}},0)=1$, which means the optimal transmit beamforming and random beamforming achieve the same PKG performance in the i.i.d. fading channels.
%Since $f_{l}(K,L,\rho) \geq 1$ and the equality holds in the case of $\rho=0$, $\boldsymbol{w}_{opt}^T \boldsymbol{R}_{S} \boldsymbol{w}_{opt}^* \geq P_{A}$ and achieve larger KGR gain with the increase of correlation and the number of antennas. 
In addition, it can be observed that both the upper and lower bounds converge to $P_\mathrm{A} (\frac{1+\rho}{1-\rho})^2$ as ${N_{\mathrm{H}}^{t}}\rightarrow \infty$ and ${N_{\mathrm{V}}^{t}}\rightarrow \infty$. This means when the BS is equipped with a large amount of antennas, the KGR depends only on the correlations among the antennas of the BS for a given power. 
Also, the KGR increases monotonically with the correlation coefficient $\rho$. 

\subsection{KGR under Different Assumptions of Channel Model at RIS}
	In Lemma 1, the KGR is proportional to $\boldsymbol{v}^H \tilde{\boldsymbol{R}}_{I}\boldsymbol{v}$. 
	Under the assumption of the i.i.d. channel model adopting in existing works, the spatial correlation matrix is $\tilde{\boldsymbol{R}}_{I}=\boldsymbol{I}$. By employing the random reflection, the expectation of the objective function of $\mathcal{P}_{2}$ is $\mathbb{E}\{\widetilde{\boldsymbol{v}}^H \tilde{\boldsymbol{R}}_{I} \widetilde{\boldsymbol{v}}\} = N,$
%	\begin{align}
%		\mathbb{E}\{\widetilde{\boldsymbol{v}}^H \tilde{\boldsymbol{R}}_{I} \widetilde{\boldsymbol{v}}\} = N,
%	\end{align}
	where each phase in $\widetilde{\boldsymbol{v}}$ can be drawn from the uniform distribution, i.e., $\widetilde{\theta}_{i} \sim U[0,2\pi),i\in \{1,\cdots,N\}$. 
	
	Taking the spatial correlation model into account, the maximum value of $\boldsymbol{v}^H \tilde{\boldsymbol{R}}_{I}\boldsymbol{v}$ is $||\boldsymbol{R_{I} }||_{F}^2 $, where $||\cdot||_{F}^2$ denotes the Frobenius norm. 
	To characterize the impact of spatial correlation on RIS, we have the following lemma.
	%the physically feasible spatial correlation model at RIS. 
	\begin{lemma}
		(Proposition 1 in \cite{Rayleigh-fading}) In isotropic scattering environments, the spatial correlation of RIS is expressed as 
		\begin{align}
			[\boldsymbol{R}_{I}]_{n, m}=\operatorname{sinc} \frac{2\left\|\boldsymbol{u}_{n}-\boldsymbol{u}_{m}\right\|}{\lambda}, \ \forall n,m \in  \{1,\cdots,N\}, \label{R_I}
		\end{align}
		%where the location of the $n$-th element is $ \boldsymbol{u}_{n} = [0,\ i(n)d_{H},\ j(n)d_{V} ]^T$ and $ i(n)= \bmod \left(n-1, N_{\mathrm{H}}\right) $ and $j(n)=\left\lfloor(n-1) / N_{\mathrm{H}}\right\rfloor$ denote horizontal and vertical indices of element $n$, respectively. 
		%the sinc function is $\operatorname{sinc}(x) = \operatorname{sinc}(\pi x)/(\pi x) $.
		where $\left\|\boldsymbol{u}_{n}-\boldsymbol{u}_{m}\right\|$ denotes the distance between $n$-th RIS element and $m$-th RIS element, $\lambda$ is the wavelength. 
	\end{lemma}
	
	%\emph{Remark 1:} 
	%Then, the element in  $\tilde{\boldsymbol{R}}_{I} $ is given by $ [\tilde{\boldsymbol{R}}_{I}]_{i,j}  = \left|\operatorname{sinc} \frac{2\left\|\boldsymbol{u}_{n}-\boldsymbol{u}_{m}\right\|}{\lambda} \right|^2$. 

	Since the sinc function $\operatorname{sinc}(x) = \operatorname{sinc}(\pi x)/(\pi x)$ is monotonically decreasing in interval $[0,1)$, 
	the entries in $ \tilde{\boldsymbol{R}}_{I} $ is larger as the inter-element spacing becomes smaller, when the elements distance fulfill $\left\|\boldsymbol{u}_{n}-\boldsymbol{u}_{m}\right\| \leq \frac{\lambda}{2} $. 
	Moreover, the optimal value of $\boldsymbol{v}^H\tilde{\boldsymbol{R}}_{I}\boldsymbol{v}$ satisfies $|| \boldsymbol{R}_{I}||_{F}^2 > N$ because the correlation between the elements always exists in practical RIS systems \cite{Rayleigh-fading}. Hence, the KGR performance of the proposed reflective beamforming is better than the counterpart adopting the assumption of the i.i.d. channel model. 
%\subsection{Special case: No correlation}
%\emph{Remark 2:} 
%It is also noticed that for the special case RIS is uncorrelated, the key rate is given by 
%\begin{align}
%	R_{SK} = 
%	\log\left(1+ \frac{ (\frac{P}{\sigma^2})^2 \lambda_{m}^2 (\beta_{r} N + \beta_d )^2}{2\frac{P}{\sigma^2} \lambda_{m} (\beta_{r} N + \beta_d )  + 1} \right)
%\end{align}
%which is independent of the reflective beamforming vector. 
%With the number of RIS elements increase, the key generation rate $ \mathcal{I}\left(\tilde{ h}_{a} ; \tilde{ h}_{b} \right) \rightarrow \infty $.

%At the same time, as the transmit power $ P_A \rightarrow \infty $, the rate is 
%\begin{align}
%	R_{SK}  \rightarrow 
%	\log\left(1+ P_B (\beta_{r} N + \beta_d ) /\sigma^2\right)
%\end{align}
%which is a logarithmic function of the SNR and the number of RIS elements. 

%% 
\section{Simulation Results}
In this section, we evaluate the performance of the proposed method with the aid of numerical simulations. 
We assume that Alice, Bob, and RIS are located at (0\ m, 0\ m), (70\ m, 0\ m), and (50\ m, 10\ m), respectively\footnote{Since the eavesdropping channels are independent of the legitimate channels in this paper, the exact location of Eve and the number of Eve's antennas has no impact on the KGR.} \cite{Globecom}. Alice is equipped with a UPA antenna. The RIS is a uniform rectangular array (URA) with $N_{\mathrm{H}}^r $ elements per row and $ N_{\mathrm{V}}^r$ elements per column. 
%We consider a correlated Rayleigh fading model as shown in Sec.II-A. 
The large-scale path loss $ \beta_{ba} =\sqrt{\zeta_{0} d_{ba}^{-\alpha_{ba}} } $, where $d_{ba}$, $\zeta_{0}$, and $\alpha_{ba}$ are the distance, path loss at 1 m, and the path loss exponent, respectively. The transmit power are $P_\mathrm{A}=P_\mathrm{B}=P$ \cite{li2021maximizing}, \cite{JiOTP}, and the simulation settings are $\alpha_{ba} = 4 $, $\alpha_{br} = \alpha_{ar} = 2 $, $\zeta_{0}=-30$ dB, and $\sigma^2 = -80$\ dBm \cite{Globecom}. 
\begin{figure}
	\centering
	\includegraphics[width=3.1in]{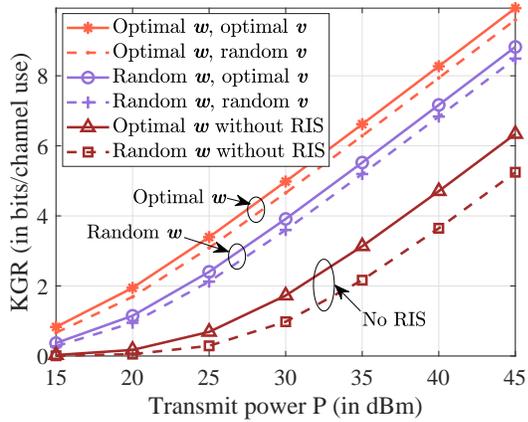}
	\caption{The KGR achieved by different beamforming settings when $M=16$, $\rho=0.3$, $N=64$, and RIS element spacing is $\lambda/2$.}
	\label{KGR_P}
\end{figure}
\subsection{Optimality of the Proposed Method}
In Fig. \ref{KGR_P}, the KGR versus the transmit power, $P$, is plotted for different transmit and reflective beamforming settings.
%\blue{ The number of antennas at BS and elements at RIS is $M= K \times L = 16$ and $N = N_{H} \times N_{V} = 25$, respectively. 
First, we observe that the KGR at all settings increases with the transmit power, since the negative impacts of noises are reduced. For comparison, the benchmarks are random beamforming based on the i.i.d. channel model and the case without RIS. 
It is noted that the proposed optimal design outperforms these benchmarks. 
Specifically, when $P\geq20$ dBm, the optimal setting achieves about $5$ dB and $11$ dB transmit power gain compared to the beamforming scheme under the i.i.d. channel assumption and the optimal transmit beamforming without RIS, respectively. 
This is because when correlations exist between the BS antennas and the RIS elements, the i.i.d. model fails in capturing this characteristic which degrades the KGR performance. 
In contrast, the proposed scheme can effectively exploit the properties of the channels to perform precise beamforming. 
Finally, the KGR gain of optimizing $\boldsymbol{w}$ is larger than that of optimizing $\boldsymbol{v}$. Indeed, optimizing $\boldsymbol{w}$ is more effective than that of $\boldsymbol{v}$ in combating the noises in $R_{\mathrm{SK}}$, and this aligns with the analysis in (\ref{eq:whole}).

% since in this simulation,  $\boldsymbol{w}$ realizes the optimal power allocation. 

\begin{figure}
	\centering
	\includegraphics[width=3.1in]{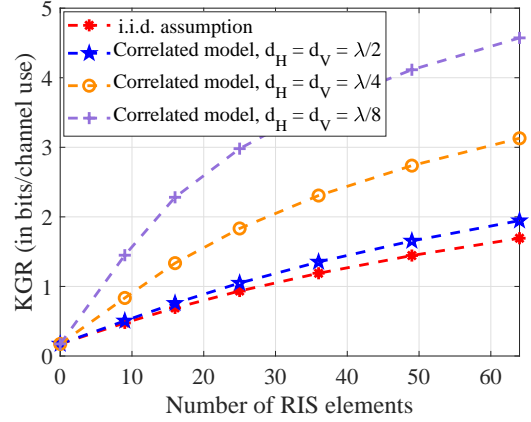}
	\caption{The KGR achieved for different $N$ with $N_{\mathrm{H}}^r=N_{\mathrm{V}}^r$, when $P=20$ dBm, $M=16$, and $\rho = 0.3$.}
	\label{KGR_N}
\end{figure}

\subsection{The Impact of RIS Elements Number and Size}
Fig. \ref{KGR_N} shows the KGR of different spatial correlations at RIS versus the number of RIS elements $N$. 
It is observed that the KGR of all of these cases increases with the number of RIS elements. As more elements are placed, more  electromagnetic signals are reflected by the RIS to realize better KGR performance.
%the proposed design becomes more flexible to create pencil-like energy focusing beams at the RIS to realize better KGR performance. 
Moreover, we notice that with the elements spacing becomes smaller, the KGR increases significantly. 
This is because with smaller elements spacing, the values of the spatial correlation matrix $\boldsymbol{R}_{I}$ are larger, contributing to a higher KGR. 
%These results are consistent with the analysis in Sec. III-B. Also, 
Also, it is found that even with $\lambda/2$ RIS element spacing, the KGR of the proposed method is still slightly superior than that adopting the i.i.d. assumption. In fact, the correlation among the RIS elements is weak in $\lambda/2$ spacing, although it always exists if $N_{\mathrm{H}}^{r} > 1$ and $N_{\mathrm{V}}^{r} > 1 $, which can be exploited by the proposed method.

%Since an RIS is envisioned to be implemented with element spacing $ d \in [\lambda/8, \lambda/4] $ \cite{ozdogan2019intelligent}, the performance gain of RIS spatial correlation on KGR should not be ignored.

\subsection{The Impact of BS Antennas Number and Correlation} 
\begin{figure}
	\centering
	\includegraphics[width=3.1in]{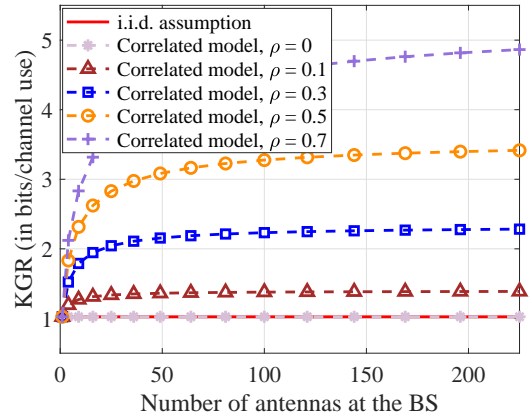}
	\caption{The KGR versus the number of antennas $M$ with $N_{\mathrm{H}}^{t}=N_{\mathrm{V}}^{t}$, when $N=64$, $P=20$ dBm, and RIS element spacing is $\lambda/2$.}
	\label{KGR_M} 
\end{figure}
Fig. \ref{KGR_M} shows the KGR versus the number of the antennas at the BS. As can be observed, the KGR of the design method based on the i.i.d. fading model is identical to that of the proposed design when $\rho = 0$, which is independent of antenna number at the BS. For the cases of $\rho>0$, the proposed method can achieve higher KGR gain, since the upper and lower bounds of the KGR both increase with the spatial correlation between antennas. 
Moreover, with the number of antennas increases, the KGR increases with diminishing returns. This is due to the channel hardening and the limited transmit power at the BS. 
%by the maximum transmit power constraint  
%The greater correlation coefficient $\rho$ contributes to higher convergence value, which is consistent with Lemma 3. 

%% 
\section{Conclusion}
In this paper, we introduced a novel transmit and passive beamforming based RIS-assisted PKG method in multi-antennas spatially correlated channels. We formulated the optimization problem and obtained a globally optimal solution to maximize the KGR. We compared the KGR performance under the assumptions of the  i.i.d. channel model and the spatially correlated channel model. We found that in spatially correlated channels, the proposed beamforming design achieves higher PKG than that under the i.i.d. channel model assumption. 
Simulation results confirmed the performance of the proposed method and the analysis of the spatial correlation.

\appendices
\section{Covariance Calculation} % (fold)
\label{sec:proof_of_thm1}
First, we calculate the covariance of channel $\hat{ h}_{a}$ as 
\begin{align}
	\mathcal{R}_{a,a} 
%	 &=\mathbb{E}\left\{ \tilde{ h}_{a} \tilde{ h}_{a}^H \right\} \\ 
%	&= P_{B}\mathbb{E}\{ \boldsymbol{w}^T (\boldsymbol{G}_{ra} \boldsymbol{\Phi} \boldsymbol{h}_{br} + \boldsymbol{h}_{ba}  )(\boldsymbol{G}_{ra} \boldsymbol{\Phi} \boldsymbol{h}_{br} + \boldsymbol{h}_{ba}  )^H   \boldsymbol{w}^* \} + P_{A}\sigma_{a}^{ 2 }   \\
%	&= \mathbb{E}\{ \boldsymbol{w}^T \boldsymbol{G} \boldsymbol{\Phi} \boldsymbol{h}_{br} \boldsymbol{h}_{br}^H \boldsymbol{\Phi}^H \boldsymbol{G}^H \boldsymbol{w}^* +    \boldsymbol{w}^T \boldsymbol{h}_{ba}  \boldsymbol{h}_{ba}^H \boldsymbol{w}^*   \} + P_{A}\sigma_{a}^{ 2 }  \\
	& =P_\mathrm{B} \boldsymbol{w}^T \mathbb{E}\{  \boldsymbol{G}_{ra} \boldsymbol{\Phi} \boldsymbol{h}_{br} \boldsymbol{h}_{br}^H \boldsymbol{\Phi}^H \boldsymbol{G}_{ra}^H  \} \boldsymbol{w}^*    \notag \\ & \quad +    P_\mathrm{B} \boldsymbol{w}^T \mathbb{E}\{\boldsymbol{h}_{ba}  \boldsymbol{h}_{ba}^H \} \boldsymbol{w}^* +  ||\boldsymbol{w}||^{2} \sigma_{a}^{ 2 }. \label{Ra}
\end{align}
Assuming the BS has obtained the $\boldsymbol{R}_{S}$ by employing \cite{neumann2018covariance}, we can calculate the first term in (\ref{Ra}) as
\begin{align}
	& \quad \mathbb{E}\{  \boldsymbol{G}_{ra} \boldsymbol{\Phi} \boldsymbol{h}_{br} \boldsymbol{h}_{br}^H \boldsymbol{\Phi}^H \boldsymbol{G}_{ra}^H  \} \\
%	&= \beta_{ r }  \boldsymbol{R}_{s} 
%	\mathbb{E}\{    \tilde{ \boldsymbol{H}}_{ar} \boldsymbol{R}_{I}^{\frac{1}{2}}     \boldsymbol{\Phi}  \boldsymbol{R}_{I}^{\frac{1}{2}} \tilde{ \boldsymbol{h}}_{br} \tilde{ \boldsymbol{h}}_{br} ^H \boldsymbol{R}_{I}^{\frac{1}{2}}  \boldsymbol{\Phi}^H  \boldsymbol{R}_{I}^{\frac{1}{2}}  \tilde{ \boldsymbol{H}}_{ar}^H      \}  \\
%	& = \frac{\beta_{ r }}{ M }  \boldsymbol{R}_{s}  
%	\mathbb{E}\{\tilde{ \boldsymbol{h}}_{br} ^H \boldsymbol{R}_{I}^{\frac{1}{2}}  \boldsymbol{\Phi}^H  \boldsymbol{R}_{I}^{\frac{1}{2}}  \tilde{ \boldsymbol{H}}_{ar}^H \tilde{ \boldsymbol{H}}_{ar} \boldsymbol{R}_{I}^{\frac{1}{2}}     \boldsymbol{\Phi}  \boldsymbol{R}_{I}^{\frac{1}{2}} \tilde{ \boldsymbol{h}}_{br}          \}  \\
%	& =   \beta_{ r }  \boldsymbol{R}_{s}  
%	\mathbb{E}\{\tilde{ \boldsymbol{h}}_{br} ^H \boldsymbol{R}_{I}^{\frac{1}{2}}  \boldsymbol{\Phi}^H  \boldsymbol{R}_{I}^{\frac{1}{2}}   \boldsymbol{R}_{I}^{\frac{1}{2}}     \boldsymbol{\Phi}  \boldsymbol{R}_{I}^{\frac{1}{2}} \tilde{ \boldsymbol{h}}_{br}          \}   \\ 
	& = \beta_{ r }  \boldsymbol{R}_{S}  
	\mathbb{E}\{  \text{vec}\{\tilde{ \boldsymbol{h}}_{br} ^H \boldsymbol{R}_{I}^{\frac{1}{2}}  \boldsymbol{\Phi}^H  \boldsymbol{R}_{I}^{\frac{1}{2}} \}^H    \text{vec}\{\tilde{ \boldsymbol{h}}_{br} ^H \boldsymbol{R}_{I}^{\frac{1}{2}}  \boldsymbol{\Phi}^H  \boldsymbol{R}_{I}^{\frac{1}{2}} \}        \}  \\   
	& = \beta_{ r }  \boldsymbol{R}_{S}  
	\mathbb{E}\{   \boldsymbol{v}^H   ( (\boldsymbol{R}_{I}^{\frac{1}{2}} )^T \odot  (\boldsymbol{R}_{I}^{\frac{1}{2}} ) )^H (\tilde{ \boldsymbol{h}}_{br} ^* \otimes \boldsymbol{I}_{N}   )      \notag \\ 
	& \quad     \times(\tilde{ \boldsymbol{h}}_{br} ^T \otimes \boldsymbol{I}_{N}   )  ( (\boldsymbol{R}_{I}^{\frac{1}{2}} )^T \odot  (\boldsymbol{R}_{I}^{\frac{1}{2}} ) )  \boldsymbol{v}     \}  \\  
	& = \beta_{ r }  \boldsymbol{R}_{S}  
	  \boldsymbol{v}^H   ( (\boldsymbol{R}_{I}^{\frac{1}{2}} )^T \odot  (\boldsymbol{R}_{I}^{\frac{1}{2}} ) )^H       ( (\boldsymbol{R}_{I}^{\frac{1}{2}} )^T \odot  (\boldsymbol{R}_{I}^{\frac{1}{2}} ) )  \boldsymbol{v}      \\ 
	  & = \beta_{ r }  \boldsymbol{R}_{S}  
	  \boldsymbol{v}^H   ( \boldsymbol{R}_{I} ^{ T } \circ   \boldsymbol{R}_{I})           \boldsymbol{v}, %= \beta_{ r }  \boldsymbol{R}_{s}  
	  %\boldsymbol{v}^H   ( \boldsymbol{R}_{I} ^{ T } \circ  \tilde{\boldsymbol{R}}_{I} ) \boldsymbol{v}
\end{align}
%where $ \circ $ is Hadamard product,  $\odot $ is Khatri-Rao product, and $\otimes $ is Kronecker product. 
where $\otimes$ and $\odot$ denote the Kronecker product and Khatri-Rao product, respectively,   
%(a) follows from $ \text{vec}( \A \B \C ) = (\C^T \otimes \A) \B$ and $ \text{vec}( \A \diag(\d) \C ) = (\C^T \odot \A) \d$,
%where 
$\text{vec}(\boldsymbol{X})$ denotes the vectorization of a matrix, and 
%At the same time, we can obtain that 
%\begin{align}
%	\boldsymbol{w}^T \mathbb{E}\{\boldsymbol{h}_{ba}  \boldsymbol{h}_{ba}^H \} \boldsymbol{w}^{ * }= \beta_{d} \boldsymbol{w}^T \boldsymbol{R}_{s} \boldsymbol{w}^{ * }
%\end{align}
$\circ$ denotes Hadamard product. 
Then, the second term in (\ref{Ra}) is calculated as $\mathbb{E}\{\boldsymbol{h}_{ba}  \boldsymbol{h}_{ba}^H \} = \beta_{ ba } \boldsymbol{R}_{S}$. 
%Thus, the covariance is obtained as
%\begin{align}
%%	\mathcal{R}_{a} = \beta_{ r } \boldsymbol{w}^T  \boldsymbol{R}_{s}  \boldsymbol{w}^{ * }
%%	\boldsymbol{v}^H   ( \boldsymbol{R}_{I} ^{ T } \circ   \boldsymbol{R}_{I})           \boldsymbol{v} + \sigma_{ a } ^ { 2 }
%\mathcal{R}_{a,a} = P_B\boldsymbol{ w }^{ T } \boldsymbol{R}_S \boldsymbol{w}^*(\beta_{ r }  \boldsymbol{v}^H (\boldsymbol{R}_{I} ^{ T } \circ   \boldsymbol{R}_{I})\boldsymbol{v} + \beta_{ ba } ) + P_{A}\sigma^2.
%\end{align}
%Similarly, the covariance of  channel $\tilde{ h}_{b}$ is expressed as 
%\begin{align}
%	\mathcal{R}_{b} =\boldsymbol{ w }^{ T } \boldsymbol{R}_s \boldsymbol{w}^*(\beta_{ r }  \boldsymbol{v}^H (\boldsymbol{R}_{I} ^{ T } \circ   \boldsymbol{R}_{I})\boldsymbol{v} + \beta_{ d } ) +  \sigma_{ b }^{ 2 }
%%	&= \mathcal{R}_{ab} + \sigma_{ b } ^ { 2 } \\
%%	& = \beta_{ r } \boldsymbol{w}^T  \boldsymbol{R}_{s}  \boldsymbol{w}^{ * }
%%	\boldsymbol{v}^H   ( \boldsymbol{R}_{I} ^{ T } \circ   \boldsymbol{R}_{I})           \boldsymbol{v} + \sigma_{ b } ^ { 2 } 
%\end{align}
Other covariances can be calculated similarly and are omitted here.

\section{Proof of Lemma~\ref{lemma1}} % (fold)
\label{sec:proof_B}
We first denote $\boldsymbol{w}=\sqrt{P}\boldsymbol{w}_{0}$ with $||\boldsymbol{w}_{0}||^2=1$. Since $ \frac{d R_{\mathrm{SK}}}{dP} \ge 0$, the optimal $P$ is $P_\mathrm{A}$. Then, consider the function  
%$f(x) = \left((P_{B}x + P_{A}\sigma^2)(x + \sigma^2 )\right)/\left((P_A\sigma^{ 2 } + P_B\sigma^2)x   + P_A\sigma^4\right),$
\begin{align}
	f(x) = \frac{(P_\mathrm{B}x + P_\mathrm{A}\sigma^2)(x + \sigma^2 )}{(P_\mathrm{A}\sigma^{ 2 } + P_\mathrm{B}\sigma^2)x   + P_\mathrm{A}\sigma^4},
\end{align}
that is monotonically increasing for $x > 0$, since 
\begin{align}
	\frac{df(x)}{dx} = \frac{P_\mathrm{B}}{\sigma^2} \frac{(P_\mathrm{A} + P_\mathrm{B})x^2 + 2P_\mathrm{A} \sigma^2 x}{( (P_\mathrm{A}+ P_\mathrm{B})x +P_\mathrm{A} \sigma^2)^2 }>0 .
\end{align}
Denote $x = \boldsymbol{ w }^{ T } \boldsymbol{R}_s \boldsymbol{w}^*(\beta_{ r }  \boldsymbol{v}^H \tilde{\boldsymbol{R}}_{I}\boldsymbol{v} + \beta_{ ba } )$
and the objective function is $R_{\mathrm{SK}} = \log_{2} f(x)$. This completes the proof. 
%\begin{align}
%	\mathcal{I}\left(\tilde{ h}_{a} ; \tilde{ h}_{b}\right) = \log f(x) 
%\end{align}
%Consequently, the mutual information $ \mathcal{I}\left(\tilde{ h}_{a} ; \tilde{ h}_{b}\right)$ increases monotonously with $x$, 

%We first introduce the following lemma:
%\begin{lemma}\label{la:in}
%	The fractional function
%	\begin{align}
%		f(x) = \frac{(P_{B}x + P_{A}\sigma^2)(x + \sigma^2 )}{(P_A\sigma^{ 2 } + P_B\sigma^2)x   + P_A\sigma^4},
%	\end{align}
%	is monotonically increasing for $x > 0$.
%\end{lemma}
%
%We can prove this lemma by derivating $f(x)$ as  
%\begin{align}
%	\frac{df(x)}{dx} = \frac{P_{B}}{\sigma^2} \frac{(P_{A} + P_{B})x^2 + 2P_{A} \sigma^2 x}{( (P_{A} + P_{B})x +P_{A} \sigma^2)^2 } .
%\end{align}
%For $x > 0$, the derivative  $\frac{df(x)}{dx} \ge 0$, proving the  monotonically increasing property of $f(x)$.
%Then, denote $x = \boldsymbol{ w }^{ T } \boldsymbol{R}_s \boldsymbol{w}^*(\beta_{ r }  \boldsymbol{v}^H \tilde{\boldsymbol{R}}_{I}\boldsymbol{v} + \beta_{ ba } )$
% and the objective function can be expressed as $R_{SK} = \log f(x)$,
%%\begin{align}
%%	\mathcal{I}\left(\tilde{ h}_{a} ; \tilde{ h}_{b}\right) = \log f(x) 
%%\end{align}
%%Consequently, the mutual information $ \mathcal{I}\left(\tilde{ h}_{a} ; \tilde{ h}_{b}\right)$ increases monotonously with $x$, 
%which completes the proof. 

\section{Proof of  Lemma ~3} % (fold)
\label{section:proof-lemma3}
	Since $\boldsymbol{w}_{\mathrm{opt}}^T \boldsymbol{R}_{S} \boldsymbol{w}_{\mathrm{opt}}^* = P_\mathrm{A}\lambda_{\mathrm{max}}(\boldsymbol{R}_{S})$, we have \cite{choi2014bounds}
	\begin{align}
		\lambda_{\mathrm{max}}( \boldsymbol{R}_{S}) &\approx 
%		 \lambda_{\mathrm{max}}(\boldsymbol{R}_{h} \otimes \boldsymbol{R}_{v}) \notag \\
%		&=\lambda_{\mathrm{max}}(\left(\mathbf{U}_{h} \boldsymbol{\Lambda}_{h} \mathbf{U}_{h}^{H}\right) \otimes\left(\mathbf{U}_{v} \boldsymbol{\Lambda}_{v} \mathbf{U}_{v}^{H}\right) ) \\
%		&=\lambda_{\mathrm{max}}(\left(\mathbf{U}_{h} \otimes \mathbf{U}_{v}\right)\left(\boldsymbol{\Lambda}_{h} \otimes \boldsymbol{\Lambda}_{v}\right)\left(\mathbf{U}_{h} \otimes \mathbf{U}_{v}\right)^{H} ) \\
%		& = 
		\lambda_{\mathrm{max}}( \boldsymbol{R}_{h} )\lambda_{\mathrm{max}}( \boldsymbol{R}_{v} ),
	\end{align} 
where $\lambda_{\mathrm{max}}(\cdot)$ returns the maximun eigenvalue of the input matrix. 
%	Then, for a Toeplitz matrix $ \boldsymbol{R}_{l},l=h,v$, we extend it to a circulant matrix $ \boldsymbol{R}_{c}^{l} $ as 
Then, we extend $ \boldsymbol{R}_{l},l\in \{h,v\}$ to a circulant matrix
	\begin{align}
		\boldsymbol{R}_{l}^{c}=\left[\begin{array}{ccccccccc}
			1 & \rho & \cdots & \rho^{N_{l}^{t}-1} & \rho^{N_{l}^{t}-2}   & \cdots & \rho^2 & \rho \\
			\rho & 1 & \cdots & \rho^{N_{l}^{t}-2} & \rho^{N_{l}^{t}-1} & \cdots & \rho^3 &\rho^{2} \\
			\vdots & \vdots & \cdots & & \ddots & & &\vdots \\
			\rho & \rho^{2} & \cdots & \rho^{N_{l}^{t}-2} & \rho^{N_{l}^{t}-3}  & \cdots &\rho & 1
		\end{array}\right], \notag
	\end{align}
where $N_{l}^{t}\in \{N_{\mathrm{H}}^{t},N_{\mathrm{V}}^{t}\}$ and $\boldsymbol{R}_{l}$ is in the first $N_{l}^{t}$ rows and $N_{l}^{t}$ columns. 
According to the Cauchy Interlace Theorem \cite{hwang2004cauchy}, 
%we have 
%$ \lambda_{\mathrm{max}}(\boldsymbol{R}_{l}) \leq \lambda_{\mathrm{max}}(\boldsymbol{R}_{l}^{c}) $ and  
%	\begin{align}
%		\lambda_{\mathrm{max}}(\boldsymbol{R}_{l}^{c}) = \boldsymbol{u}_{\mathrm{max}}^H \boldsymbol{R}_{l}^{c} \boldsymbol{u}_{\mathrm{max}} = \frac{(1+\rho)(1-\rho^{N_{l}-1})}{ 1-\rho },
%	\end{align}
\begin{align}
	\lambda_{\mathrm{max}}(\boldsymbol{R}_{l}) \leq \frac{1}{2N_l^{t} -1} \boldsymbol{u}_{\mathrm{max}}^H \boldsymbol{R}_{l}^{c} \boldsymbol{u}_{\mathrm{max}} = \frac{(1+\rho)(1-\rho^{N_{l}^{t}-1})}{ 1-\rho }, \notag
\end{align}
where $\boldsymbol{u}_{\mathrm{max}}=[1,\cdots,1]^T$.
%is the eigenvector corresponding to the largest eigenvalue of $\boldsymbol{R}_{c}^{l}$. 
%Since the eigenvectors of $\boldsymbol{R}_{c}$ is DFT matrix, we can take the first column of the DFT matrix, $ \boldsymbol{u}_{m}$ to get the maximun eigenvalue of $ \boldsymbol{R}_{c}$ as follows
%To establish a lower bound, 
%	according the Rayleigh quotient inequality, for any vector  $\boldsymbol{x}$ that satisfies $||\boldsymbol{x}||_{2}  = 1$, we have 
%	\begin{align}
%		\lambda_{m}(\boldsymbol{R}_{l}) \geq \boldsymbol{x}^H \boldsymbol{R}_{l} \boldsymbol{x}
%	\end{align}
	%	Considering the elements of $\boldsymbol{R}_{l}$ are positive values, 
Let $ \boldsymbol{x}_{0} = \frac{1}{\sqrt{N_{l}^{t}}} [1,1,\cdots,1]^T$ and 
%we can obtain
	\begin{align}
		\lambda_{\mathrm{max}}(\boldsymbol{R}_{l}) \geq \boldsymbol{x}^H_{0} \boldsymbol{R}_{l} \boldsymbol{x}_{0} 
%		& = \left( 2\sum_{l=0}^{M -1 } \sum_{k=0}^{l} \rho^k - M \right)\frac{1}{M} \notag \\
		=\frac{1+\rho}{1 - \rho} - \frac{2\rho(1- \rho^{N_{l}^{t}})}{N_{l}^{t} (1-\rho)^2}.
	\end{align}

\bibliographystyle{IEEEtran}
\bibliography{IEEEabrv,mmm}

\begin{thebibliography}{10}
\providecommand{\url}[1]{#1}
\csname url@samestyle\endcsname
\providecommand{\newblock}{\relax}
\providecommand{\bibinfo}[2]{#2}
\providecommand{\BIBentrySTDinterwordspacing}{\spaceskip=0pt\relax}
\providecommand{\BIBentryALTinterwordstretchfactor}{4}
\providecommand{\BIBentryALTinterwordspacing}{\spaceskip=\fontdimen2\font plus
\BIBentryALTinterwordstretchfactor\fontdimen3\font minus
  \fontdimen4\font\relax}
\providecommand{\BIBforeignlanguage}[2]{{%
\expandafter\ifx\csname l@#1\endcsname\relax
\typeout{** WARNING: IEEEtran.bst: No hyphenation pattern has been}%
\typeout{** loaded for the language `#1'. Using the pattern for}%
\typeout{** the default language instead.}%
\else
\language=\csname l@#1\endcsname
\fi
#2}}
\providecommand{\BIBdecl}{\relax}
\BIBdecl

\bibitem{li2021maximizing}
G.~Li, C.~Sun, W.~Xu, M.~Di~Renzo, and A.~Hu, ``On maximizing the sum secret
  key rate for reconfigurable intelligent surface-assisted multiuser systems,''
  \emph{IEEE Trans. Inf. Forensics Security}, vol.~17, pp. 211--225, 2021.

\bibitem{JiOTP}
Z.~Ji, P.~L. Yeoh, G.~Chen, C.~Pan, Y.~Zhang, Z.~He \emph{et~al.}, ``Random
  shifting intelligent reflecting surface for {OTP} encrypted data
  transmission,'' \emph{IEEE Wireless Commun. Lett.}, vol.~10, no.~6, pp.
  1192--1196, 2021.

\bibitem{Low-entropy}
P.~Staat, H.~Elders-Boll, M.~Heinrichs, R.~Kronberger, C.~Zenger, and C.~Paar,
  ``{Intelligent reflecting surface-assisted wireless key generation for
  low-entropy environments},'' in \emph{Proc. IEEE Int. Symp. Person. Indoor
  Mobile Radio Commun. (PIMRC)}, Virtual, Sep. 2021, pp. 1--7.

\bibitem{Ji}
Z.~{Ji}, P.~L. {Yeoh}, D.~{Zhang}, G.~{Chen}, Y.~{Zhang}, Z.~{He}
  \emph{et~al.}, ``Secret key generation for intelligent reflecting surface
  assisted wireless communication networks,'' \emph{IEEE Trans. Veh. Technol.},
  pp. 1--1, 2020.

\bibitem{SPletter}
X.~{Lu}, J.~{Lei}, Y.~{Shi}, and W.~{Li}, ``Intelligent reflecting surface
  assisted secret key generation,'' \emph{IEEE Signal Process. Lett.}, pp.
  1--1, 2021.

\bibitem{Rayleigh-fading}
E.~Björnson and L.~Sanguinetti, ``Rayleigh fading modeling and channel
  hardening for reconfigurable intelligent surfaces,'' \emph{IEEE Wireless
  Commun. Lett.}, vol.~10, no.~4, pp. 830--834, 2021.

\bibitem{li2021reconfigurable}
G.~Li, L.~Hu, P.~Staat, H.~Elders-Boll, C.~Zenger, C.~Paar \emph{et~al.},
  ``Reconfigurable intelligent surface for physical layer key generation:
  Constructive or destructive?'' \emph{accepted by IEEE Wireless Commun.},
  2022.

\bibitem{neumann2018covariance}
D.~Neumann, M.~Joham, and W.~Utschick, ``Covariance matrix estimation in
  massive {MIMO},'' \emph{IEEE Signal Process. Lett.}, vol.~25, no.~6, pp.
  863--867, 2018.

\bibitem{choi2014bounds}
J.~Choi and D.~J. Love, ``Bounds on eigenvalues of a spatial correlation
  matrix,'' \emph{IEEE Commun. Lett.}, vol.~18, no.~8, pp. 1391--1394, 2014.

\bibitem{Globecom}
G.~Zhou, C.~Pan, H.~Ren, K.~Wang, and A.~Nallanathan, ``Outage constrained
  transmission design for {IRS}-aided communications with imperfect cascaded
  channels,'' in \emph{IEEE Glob. Commun. Conf.(GLOBECOM)}, 2020, pp. 1--6.

\bibitem{hwang2004cauchy}
S.-G. Hwang, ``Cauchy's interlace theorem for eigenvalues of hermitian
  matrices,'' \emph{Am Math Mon}, vol. 111, no.~2, pp. 157--159, 2004.

\end{thebibliography}


\end{document}